\documentclass[onecolumn,showpacs]{revtex4}

\usepackage{graphicx}
\usepackage{dcolumn}
\usepackage{bm}

\input epsf

\begin{document}

\title{\Large  Quasi-Spherical Gravitational Collapse in higher dimension and the effect of equation of state}

\author{\bf Sanjukta Chakraborty$^1$,~Subenoy
Chakraborty$^1$\footnote{subenoyc@yahoo.co.in} and Ujjal
Debnath$^2$\footnote{ujjaldebnath@yahoo.com}}

\affiliation{$^1$Department of Mathematics, Jadavpur University,
Calcutta-32, India.\\ $^2$Department of Mathematics, Bengal
Engineering and Science University, Shibpur, Howrah-711 103,
India.\\}

\date{\today}

\begin{abstract}
 Gravitational collapse in (n+2) dimensional
quasi-spherical space-time is studied for a fluid with non
vanishing radial pressure. An exact analytic solution is obtained
(ignoring the arbitrary integration function) for the equation of
state $p_{r}=(\gamma-1)\rho.$ The singularity is studied locally
by comparing the time of formation of apparent horizon and the
central shell focusing singularity while the global nature of the
final fate of collapse is characterized by the existence of radial
null geodesic. It is revealed that the end state of collapse for D
dimension with equation of state $p=-\rho$, for (D$-$1)
dimensional dust and (D$-$2)dimension with equation of state
$p=\rho$ are identical.
\end{abstract}


\maketitle

\section{\normalsize\bf{Introduction}}

 Gravitational collapse  is an important and challenging issue in Einstein gravity, particularly
  after the formation of famous singularity theorems [1] and cosmic censorship conjecture (CCC)[2].
  Also from the perspective of black hole physics and its astrophysical implications, it is interesting to know
  the final outcome of gravitational collapse [3] in the background of general relativity. The singularity theorems only
  tells us about the generic property of space-times in classical general relativity, but it cannot
  provide us about the detailed features of the singularities whether an external observer can
  visualize the singularity or not. Moreover, the CCC is
  incomplete [4,5] in the sense that there is no formal proof of
  it in one hand and on the otherhand there are counter examples
  of it. It should be noted that choice of initial data [6] has a
  role in characterizing the final state of collapse.\\

    Though a lot of works have been done for dust collapse [7-12]
    both for spherical and quasi-spherical models [13-18] in four
    and higher dimensional space-times but there is not much
    progress in studying gravitational collapse in
    quasi-spherical space for perfect fluid or matter with
    anisotropic pressure [19]. In recent past Dadhich etal [20]
    examined the role of the equation of state in charecterizing
    the final state of spherical collapse and showed a similarity among collapsing
    processes in different dimensions with different eq. of state.
    In this work we have studied quasi-spherically symmetric
    collapse of a fluid with non-vanishing radial pressure
    (with eq. of state $p_{r}=(\gamma-1)\rho$) in higher
    dimensional space-time.

\section{\normalsize\bf{Higher Dimensional Szekeres' Model}}
The quasi-spherical Szekeres' space-time in (n+2)dimension has
the  metric ansatz (in the comoving coordinates)
\begin{equation}
ds^{2}=dt^{2}-e^{2\alpha}dr^{2}-e^{2\beta}\sum_{i=1}^{n}dx_{i}^{2}\
\end{equation}

where  $\alpha$ and $\beta$ being functions of all space-time
co-ordinates are of the form
\begin{equation}
e^{\alpha}=R'+R~\nu',~~~e^{\beta}=R~e^{\nu}~~~~
 \text{ with}~~ R=R(r,t)~~~\text{ and} ~~~\nu=\nu(r,x_{1},...,x_{n})
\end{equation}

For the  matterfield  with  anisotropic pressure i.e.,
$T_{\mu}^{\nu}=\text{diag}(\rho,-p_{r},-p_{_{T}},......,-p_{_{T}})$,
we have from  the  Einstein equations (after some manipulation)

 \begin{equation}
 \left. \begin{array}{c}
 \rho=\frac{F'}{\mu^{n}\mu'}\\\\
 p_{r}=-\frac{\dot{F}}{\mu^{n}\dot{\mu}}\\\\
\mu=e^{\beta}
 \end{array}\right\}
 \end{equation}

 \begin{equation}
  F(r,t)=\frac{n}{2}{R^{n-1}e^{(n+1)\nu}(\dot{R}^{2}-f(r))}
\end{equation}
while from conservation equation $T^{\nu}_{\mu ;~\nu}=0$ we get

\begin{equation}
\begin{array}{c}
\dot{\rho}+\dot{\alpha}(\rho+p_{r})+n\dot{\beta}(p_{_{T}}+\rho)=0\\\\
p_{r}'+n\beta'(p_{r}-p_{_{T}})=0\\\\

\text{and}~~~ \alpha_{x_{i}}(p_{r}-p_{_{T}})=
\frac{\partial}{\partial x_{i} }p_{_{T}}, ~~~i=1,2,...,n
\end{array}
\end{equation}
Now for simplicity, we assume $p_{_{T}}=0 ~~\text{and}~~
p_{r}=(\gamma-1)\rho, 0<\gamma<2$. As a result, conservation
equations simplify to
\begin{equation}
\begin{array}{c}
~~~~~~~~~~~(i)~ \rho=\rho_{00}(r,x_{1},x_{2},...,x_{n})e^{-(\gamma\alpha+n\beta)}\\\\
~~~~~~~~~~~~~~~~~~~~~~~~~~~(ii)~ \alpha_{x_{i}}=0~~~ i.e.~~~ \alpha~~~ \text{is ~~independent~~ of}~~~ x_{i}~'s\\\\
(iii)~ p_{r}=p_{00}(t,x_{1},...,x_{n})e^{-n\beta}
\end{array}
\end{equation}

Assuming for simplicity,
\begin{equation}
p_{_{T}}=0 ~~~\text{and} ~~~ p_{r}=(\gamma-1)\rho,~~~0<\gamma<2,
\end{equation}
and integrating the
   conservation equations we get (choosing the arbitrary integration functions appropriately)
\begin{equation}
\begin{array}{c}
\rho(r,t)=\rho_{0}\frac{e^{-\gamma\alpha}}{R^{n}} \\\\
\text{and}~~~\alpha=\alpha(r,t)~~~~~(i.e.~~ \alpha ~~\text{is
independent of}~~ x_{i}~'s)
\end{array}
\end{equation}

From equations (3), (4) and (7) we have the evolution equation of
$R$ as
\begin{equation}
\dot{R^{2}}=\frac{2}{n}R^{1-n}[\rho_{0}R^{1-\gamma}C(r)^{-\gamma}+g(r)]
\end{equation}
 where $g(r)$ is an arbitrary integration function and assuming
 $C(r)=\frac{R'}{R}+\nu'$ to be independent of `t'.\\

 Thus choosing $g(r)=0$,  $R$ has the explicit solution

 \begin{equation}
 t-t_{i}=\left(\frac{\sqrt{2n}}{n+\gamma}\right)\frac{C(r)^{\frac{\gamma}{2}}}{\sqrt{\rho_{0}(r)}}[r^{\frac{n+\gamma}{2}}-R^{\frac{n+\gamma}{2}}]
 \end{equation}
 The hypersurface $t=t_{s}(r)$ describing shell focusing singularity
 is characterized by $R(t_{s}(r),r)=0$ i.e.
 \begin{equation}
t_{s}(r)-t_{i}=\left(\frac{\sqrt{2n}}{n+\gamma}\right)\frac{C(r)^{\frac{\gamma}{2}}}{\sqrt{\rho_{0}(r)}}r^{\frac{n+\gamma}{2}}
 \end{equation}
Now for regularity near the central singularity (i.e., $r=0$) we
assume

\begin{equation}
\begin{array}{c}
C(r)=\sum_{j=0}^{\infty}C_{j}~r^{j}\\\\
\rho_{0}(r)=\sum_{j=0}^{\infty}\rho_{j}~r^{n+\gamma+j}\\\\
\end{array}
\end{equation}

Then from $(11)$ the time for the central shell focusing
singularity is given by
\begin{equation}\begin{array}{c}
\hspace{-3cm}t_{0}~=~~~~lim~~~t_{s}(r)=t_{i}+\frac{\sqrt{2n}}{(n+\gamma)}\frac{C_{0}^{\gamma/2}}{\sqrt{\rho_{0}}}\\
\hspace{-6cm}~r\rightarrow 0\\\\

\end{array}
\end{equation}
The apparent horizon is characterized by $\dot{R}^{2}=1$ and if
$t_{ah}$ is the time of formation of apparent horizon then
\begin{equation}
t_{ah}(r)-t_{0}=\frac{\sqrt{2n}}{(n+\gamma)}\left[\frac{C_{0}^{\gamma/2}}{\sqrt{\rho_{0}}}(B~r+O(r^{2}))-A~r^{q}(1+O(r))\right]
\end{equation}

where, $B=\frac{1}{2}\left[
\frac{\gamma~C_{1}}{C_{0}}-\frac{\rho_{1}}{\rho_{0}}\right],~
A=\left(\frac{2}{n}\right)^{q/2}\rho_{0}^{\frac{1}{n+\gamma-2}}
C_{0}^{\frac{-\gamma}{n+\gamma-2}},~q=\frac{n+\gamma}{n+\gamma-2}~(>1)$~.\\

We note that if $B<0$ then $t_{ah}(r)<t_{0}$ i.e., trapped
surface forms earlier than the formation of singularity so we can
not visualize it (not naked) but if $B>0$ then singularity
appears before the formation of apparent horizon so it is
possible to visualize the singularity (naked). It should be
mentioned that $t_{ah}>t_{0}$ is the necessary condition for the
visibility of the singularity (locally) while $t_{ah}\leq~t_{0}$
is the sufficient condition for
the formation of a black hole (globally).\\

Further, if $C_{1}=0=\rho_{1}$~ (i.e., $a=0$) and $q<2~~~ (q>1)$
then from (14) the dominant term is $A~r^{q}$ which appears in
negative sign in the expression for $(t_{ah}(r)-t_{0})$. Hence
the collapse always leads to the formation of a black hole. Now
the restriction $q<2$ leads to $n+\gamma>4$ which implies
$D(=n+2)>6$ for $\gamma=0$, $D>5$ for $\gamma=1$ and $D>4$ for
$\gamma=2$. Therefore, the validity of cosmic censorship
conjecture in `D' or higher
dimension depends on the equation of state.\\

Secondly, if the arbitrary function $g(r)$ is non-zero then the
evolution equation (9) have complicated solution of the form
$$t-t_{i}=\sqrt{\frac{2n}{g(r)}}\frac{1}{n+1}\left\{r^{\frac{n+1}{2}}
~_{2}F_{1}[a,\frac{1}{2},a+1,-\frac{C(r)^{-\gamma}r^{1-\gamma}\rho_{0}(r)}{g(r)}]
-R^{\frac{n+1}{2}}~_{2}F_{1}[a,\frac{1}{2},a+1,
-\frac{C(r)^{-\gamma}R^{1-\gamma}\rho_{0}(r)}{g(r)}]\right\},$$
~~~$0<\gamma<1$~~~~with $a=\frac{n+1}{2(1-\gamma)}$\\

and for $1<\gamma<2$
$$t-t_{i}=\sqrt{\frac{2n}{\rho_{0}(r)}}\frac{C(r)^{\gamma/2}}{n+\gamma}
\left\{r^{\frac{n+\gamma}{2}}~_{2}F_{1}[b,\frac{1}{2},b+1,
-\frac{C(r)^{\gamma}r^{\gamma-1}g(r)}{\rho_{0}(r)}]-R^{\frac{n+\gamma}{2}}
~_{2}F_{1}[b,\frac{1}{2},b+1,-\frac{C(r)^{\gamma}R^{\gamma-1}g(r)}{\rho_{0}(r)}]\right\}$$
with $b=\frac{n+\gamma}{2(\gamma-1)}.$\\

Thus proceeding as above the time difference between the form of
trapped surface and the central singularity is given by

$$t_{ah}-t_{0}=\frac{r}{(n+1)g_{0}^{3/2}}\left[_{2}F_{1}[a,\frac{1}{2},a+1,
-\frac{C_{0}^{-\gamma}\rho_{0}}{g_{0}}]g_{1}+\frac{(n+1)C_{0}^{-1-\gamma}
(\gamma~c_{1}g_{0}\rho_{0}+C_{0}g_{1}\rho_{0}-g_{0}C_{0}\rho_{1})}{(3+n-2\gamma)g_{0}}\times\right.$$
$$\left.{_{2}F_{1}[a+1,\frac{3}{2},a+2,-\frac{C_{0}^{-\gamma}\rho_{0}}{g_{0}}]
}\right]+0(r^{2})
-2\frac{(2g_{0}/n)^{\frac{n+1}{2n-2}}}{(n+1)\sqrt{g_{0}}}~_{2}F_{1}[a,\frac{1}{2},a+1,
\frac{2g_{0}}{n}^{\frac{1-\gamma}{n-1}}\frac{\rho_{0}C_{0}^{-\gamma}}{g_{0}}]~
r^{\frac{2(1-\gamma)}{n-1}+\frac{n+1}{n-1}}$$

for $0<\gamma<1$\\

and
$$t_{ah}-t_{0}=\frac{rC_{0}^{\gamma/2}}{(n+\gamma)\rho_{0}^{3/2}}\left[-~_{2}F_{1}[b,\frac{1}{2},b+1,
-\frac{C_{0}^{\gamma}g_{0}}{\rho_{0}}]\rho_{1}+\frac{\rho_{0}\gamma}{C_{0}}
~_{2}F_{1}[b,\frac{1}{2},b+1,-\frac{C_{0}^{\gamma}g_{0}}{\rho_{0}}]C_{1}-
\frac{(n+\gamma)C_{0}^{\gamma-1}}{(n+3\gamma-2)\rho_{0}}\times\right.$$
$$\left.~_{2}F_{1}[b+1,\frac{3}{2},b+2,
-\frac{C_{0}^{\gamma}g_{0}}{\rho_{0}}](\gamma~C_{1}g_{0}\rho_{0}+C_{0}g_{1}\rho_{0}-g_{0}C_{0}\rho_{1})
\right]+0(r^{2})-\frac{2}{n+\gamma}\left(\frac{2g_{0}}{n}\right)^{\frac{n+\gamma}{2n-2}}\frac{C_{0}^{\gamma/2}}{\rho_{0}}\times$$
$$~_{2}F_{1}[b,\frac{1}{2},b+1,\frac{C_{0}^{\gamma}g_{0}}{\rho_{0}}
\left(\frac{2g_{0}}{n}\right)^{\frac{\gamma-1}{n-1}}]~r^{\frac{n+\gamma}{n-1}+\frac{2(\gamma-1)}{n-1}}$$

for $1<\gamma<2$.\\

From the above time difference, it is not easy to speculate the
end state of collapse (naked singularity or black hole) as it
involves so many arbitrary (or initial) constants. However, if it
so happens that the coefficient of $r $ vanishes then the leading
order term in $r $ may depend on $\gamma$ if the following
inequations are satisfied
      $$1<\frac{2(1-\gamma)}{n-1}+\frac{n+1}{n-1}<2,~~~~~\text{for}~~~~
      0\leq~\gamma~\leq~1$$
      $$1<\frac{n+\gamma}{n-1}+\frac{2(\gamma-1)}{n-1}<2,~~~~\text{for}~~~~
      1\leq~\gamma~\leq~2$$

In that case as before we always have black hole as the final
state of collapse. The above inequalities show that the dimension
parameter $n$ depends strongly  on $\gamma$. In particular, for
$\gamma=1$ we have $n>3$ i.e., we always have black hole solutions
for six and higher dimension which is in agreement with result in
dust collapse [11]. For different equation of state the
restriction on $n$ is as follows:\\
\[
\text {TABLE-I}
\]

\begin{center}
\begin{tabular}{|c|c|c|c|c|} \hline
~$\gamma$~~~&~~$0$~~~&~~$1$~~~&~~$4/3$~~~&~~$2$
\\ \hline
~n~~~&~~$6$~~~&~~$4$~~~&~~$5$~~~&~~$7$
\\ \hline
\end{tabular}
\end{center}
The table shows the least value of $n$ for different value of the
equation of state parameter $\gamma$ in formation of black hole.\\

\section{\normalsize\bf{Geodesics and the nature of singularity}}

As before, using for simplicity $g(r)=0=f(r)$ the scale factor
$R(t,r)$ has the simple explicit solution (choosing the initial
time $t_{i}=0$)
\begin{equation}
R=\left[r^{\frac{n+\gamma}{2}}-\frac{n+\gamma}{\sqrt{2n}}~~\frac{\sqrt{\rho_{0}(r)}}
{C(r)^{\gamma/2}}~t\right]^{\frac{2}{n+\gamma}}
\end{equation}

Following the geodesic analysis of Joshi and Dwivedi [21] for TBL
model, let us introduce the functions:\\

\begin{equation}
\begin{array}{c}
X=\frac{R}{r^{m}}\\\\
 \xi=\frac{2rB'}{B}\\\\
  \eta=\frac{rQ'}{Q}\\\\
\zeta=\frac{2B^{2}}{n~r^{m(n+\gamma-2)}}\\\\
 Q=e^{-\nu}\\\\
 \Theta=\frac{1-\frac{\xi}{n+\gamma}}{r^{\frac{(m-1)(n+\gamma)}{2}}}
\end{array}
\end{equation}

with $m\geq~1$ and $B(r)=\frac{\sqrt{\rho_{00}(r)}}{C(r)^{\gamma/2}}$.\\

In this section the visibility or non-visibility of the
singularity is characterized by examining the possibility to have
any outgoing null geodesics which are terminated in the past at
the central singularity $r=0$. Suppose this occurs at singularity
$t=t_{0}$ at which $R(t_{0},0)=0$. For convenience , we consider
the radial null geodesic, given by
\begin{equation}
\frac{dt}{dr}=e^{\alpha}=R'+R\nu'
\end{equation}
Now choosing, $U=r^{m}$, the above geodesic equation can be
written as
\begin{equation}
\frac{dR}{dU}=U(X,u)
\end{equation}

where
$U(X,u)=(1-\sqrt{\frac{\zeta}{X^{n+\gamma-2}}})\frac{H}{m}+\frac{\eta}{m}~\sqrt{\frac{\zeta}{X^{n+\gamma-4}}}$
with
$H=\frac{\xi~X}{n+\gamma}+\frac{\Theta}{X^{(n+\gamma-2)/2}}$\\\\

If we approach the singularity at $R=0$, $u=0$ along the radial
null geodesic then the limiting behaviour of the function $X$ is
given by

\begin{eqnarray}
\begin{array}{c}
X_{0}~= \\
{}%
\end{array}
\begin{array}{c}
lim~~~~~~~~~X \\
R\rightarrow 0~ u\rightarrow 0%
\end{array}
\begin{array}{c}
=~lim~~~~~~~~~ \frac{R}{u} \\
R\rightarrow 0~ u\rightarrow 0%
\end{array}
\begin{array}{c}
=~lim~~~~~~~~~ \frac{dR}{du} \\
~~~R\rightarrow 0~ u\rightarrow 0%
\end{array}
\begin{array}{c}
=~~lim~~~~~~~~ U(X,u) \\
R\rightarrow 0~ u\rightarrow 0%
\end{array}
\begin{array}{c}
=U(X_{0},0) \\
{}%
\end{array}%
\end{eqnarray}

This is a polynomial equation in $X_{0}$ with explicit form
\begin{equation}
X_{0}^{n+\gamma-1}[m-\frac{\xi_{0}}{n+\gamma}]-\Theta_{0}X_{0}^{(n+\gamma-2)/2}+
\Theta_{0}\sqrt{\zeta_{0}}+\sqrt{\zeta_{0}}X_{0}^{(n+\gamma)/2}\left[\frac{\xi_{0}}{n+\gamma}-\eta_{0}\right]=0
\end{equation}
Here the suffix $`0$' stands for the value of the variable at
$r=0$. \\

If it has at least one positive real root then it is possible to
have a radial null geodesic outgoing from the central singularity
and the singularity will be naked. The above polynomial equation
shows that whenever `$n$' occurs it is always with $\gamma$ in the
form $n+\gamma$. So if both $n$ and $\gamma$ changes keeping
$n+\gamma$ to be invariant then the final fate of the singularity
remains same i.e., if we have space-time dimension $D=n+2$ and
equation of state $p=(\gamma-1)\rho$ and we have another
space-time dimension $D=n+1$ and equation of state $p=\gamma\rho$
then the end state of
collapse will be same for both space time.\\
\newpage

$$\text{TABLE~ II}$$
\begin{center}
\begin{tabular}{|l|}
\hline\hline
~$\gamma$~~~~~$m$~~~~~$\eta_{0}$~~~~~$\xi_{0}$~~~~$\zeta_{0}$~~~~~$\Theta_{0}$%
~~~~~~~~~~~~~~~~~~~~~~~~~~~Positive roots ($X_{0}$) \\ \hline
\\
~~~~~~~~~~~~~~~~~~~~~~~~~~~~~~~~~~~~~~~~~~~~~~~~4D~~~~~~~~~~5D~~~~~~~~~~~6D~~~~~~~~~~~7D~~~~~~~~~~8D~~~~~~~~~~10D~~~~~~~~14D
\\ \hline\hline
\\
~0~~~~~~~~~~~~~~~~~~~~~~~~~~~~~~~~~~~~~~~~~~~~~~$-$~~~~~~~~~~.01~~~~~~~~~~~~.1~~~~~~~~~~~~.21~~~~~~~~~~.31~~~~~~~~~~.45~~~~~~~~~.62
\\
\\
~4/3~~~~~1~~~~-6~~~~~.05~~~.01~~~~-5~~~~~~~.03~~~~~~~~~.14~~~~~~~~~~~~.25~~~~~~~~~~~.39~~~~~~~~~~.41~~~~~~~~~~.52~~~~~~~~~.66
\\
\\
~2~~~~~~~~~~~~~~~~~~~~~~~~~~~~~~~~~~~~~~~~~~~~~~.1~~~~~~~~~~.21~~~~~~~~~~~~.31~~~~~~~~~~~.39~~~~~~~~~~.45~~~~~~~~~~.55~~~~~~~~~.67
\\  \hline
\\
~0~~~~~~~~~~~~~~~~~~~~~~~~~~~~~~~~~~~~~~~~~~~~~~0~~~~~~~~~~~$-$~~~~~~~~~~~~~~$-$~~~~~~~~~~~~$-$~~~~~~~~~~~~$-$~~~~~~~~~~~~$-$~~~~~~~~~~$-$~~~~~
\\
\\
~4/3~~~~~1~~~~0.1~~~~~1~~~~~1~~~~1~~~~~~~~13.6~~~~~~~~~$-$~~~~~~~~~~~~~~~$-$~~~~~~~~~~~~$-$~~~~~~~~~~~~$-$~~~~~~~~~~~$-$~~~~~~~~~~$-$~~~~~
\\
\\
~2~~~~~~~~~~~~~~~~~~~~~~~~~~~~~~~~~~~~~~~~~~~~~~$-$~~~~~~~~~~$-$~~~~~~~~~~~~~~~$-$~~~~~~~~~~~~$-$~~~~~~~~~~~~$-$~~~~~~~~~~~$-$~~~~~~~~~~$-$~~~~~
\\  \hline
\\
~0~~~~~~~~~~~~~~~~~~~~~~~~~~~~~~~~~~~~~~~~~~~~1.19~~~~~~~~.11~~~~~~~~~~~~~.36~~~~~~~~~~~.54~~~~~~~~~~.66~~~~~~~~~.81~~~~~~~~~~~$-$~~~~~~
\\
~~~~~~~~~~~~~~~~~~~~~~~~~~~~~~~~~~~~~~~~~~~~~~~~~~~~~~~~~~~~.96~~~~~~~~~~~~~.9~~~~~~~~~~~~.88~~~~~~~~~~.87~~~~~~~~~.86~~~~~~~~~~~$-$~~~~~
\\
\\
~4/3~~~~~4~~~~~~0~~~~~5~~~~0.1~~~4~~~~~~~~~.2~~~~~~~~~~.43~~~~~~~~~~~~~.58~~~~~~~~~~~.69~~~~~~~~~~.76~~~~~~~~~~$-$~~~~~~~~~~~$-$
\\
~~~~~~~~~~~~~~~~~~~~~~~~~~~~~~~~~~~~~~~~~~~~~~~.93~~~~~~~~~.89~~~~~~~~~~~~~.88~~~~~~~~~~~.87~~~~~~~~~~.87~~~~~~~~~~$-$~~~~~~~~~~~$-$
\\
\\
~2~~~~~~~~~~~~~~~~~~~~~~~~~~~~~~~~~~~~~~~~~~~~.36~~~~~~~~~.54~~~~~~~~~~~~~.66~~~~~~~~~~~.74~~~~~~~~~~~.81~~~~~~~~~$-$~~~~~~~~~~~$-$~~~~~
\\
~~~~~~~~~~~~~~~~~~~~~~~~~~~~~~~~~~~~~~~~~~~~~~~.9~~~~~~~~~~.88~~~~~~~~~~~~~.87~~~~~~~~~~~.87~~~~~~~~~~~.86~~~~~~~~~$-$~~~~~~~~~~~$-$~~~~~
\\  \hline
\\
~0~~~~~~~~~~~~~~~~~~~~~~~~~~~~~~~~~~~~~~~~~~~1.04~~~~~~~~~.9~~~~~~~~~~~~~~.87~~~~~~~~~~~.86~~~~~~~~~~~.86~~~~~~~~~$-$~~~~~~~~~~~$-$~~~~~
\\
\\
~4/3~~~~~1~~~~~~0~~~~~1~~~~0.1~~~1~~~~~~~~.2~~~~~~~~~~~.43~~~~~~~~~~~~~.59~~~~~~~~~~~.69~~~~~~~~~~.77~~~~~~~~~~$-$~~~~~~~~~~~$-$~~~~~
\\
~~~~~~~~~~~~~~~~~~~~~~~~~~~~~~~~~~~~~~~~~~~~~~.89~~~~~~~~~~.87~~~~~~~~~~~~~.86~~~~~~~~~~~.86~~~~~~~~~~.85~~~~~~~~~~$-$~~~~~~~~~~~$-$~~~~~
\\
\\
~2~~~~~~~~~~~~~~~~~~~~~~~~~~~~~~~~~~~~~~~~~~~.36~~~~~~~~~~.54~~~~~~~~~~~~~.66~~~~~~~~~~~.74~~~~~~~~~~~~$-$~~~~~~~~~~$-$~~~~~~~~~~~$-$~~~
\\
~~~~~~~~~~~~~~~~~~~~~~~~~~~~~~~~~~~~~~~~~~~~~~.87~~~~~~~~~~.86~~~~~~~~~~~~~.86~~~~~~~~~~~.85~~~~~~~~~~~$-$~~~~~~~~~~$-$~~~~~~~~~~~$-$~~~
\\
\\ \hline\hline
\end{tabular}
\end{center}

Due to the complicated nature of the above polynomial an exact
analytic solution is not possible for $X_{0}$. So we study the
roots by numerical methods. The above table (Table II) shows the
dependence of the nature of the roots on the variation of the
parameters involved. It is clear from the definition of the
functions in eq. (16) that the parameter $m,~ \xi_{0},~ \zeta_{0}$
are always positive while $\eta_{0}, ~\Theta_{0}$ may take both
positive and negative values. However, it is not possible to find
any definite conclusion due to the variation of the parameter
$\Theta_{0}$.

\subsection{\normalsize\bf{Strength of the naked singularity}}

The strength of a singularity (Tipler[22]) is characterized by
examining the state of a body that falls in it. If all objects
that fall into a singularity are destroyed by crushing or tidally
stretching to zero volume then the singularity is called
gravitationally strong otherwise it is called weak singularity. A
precise mathematical definition was given by Clarke and Krolak
[23]. According to them, a sufficient condition for a strong
curvature singularity is that, for at least one non space-like
geodesic with affine parameter $\lambda\rightarrow~0$ on approach
to the singularity, we must have
\begin{equation}
\begin{array}{c}
lim \\
\lambda \rightarrow 0 \\
\end{array}%
\begin{array}{c}
\lambda ^{2}R_{ij}K^{i}K^{j}>0 \\
\\
\end{array}%
\end{equation}%

with $K^{i}=\frac{dx^{i}}{d\lambda}$, the tangent vector to the
radial null geodesic. For future-directed radial null geodesics
that originate from the naked singularity the above limit can be
written as (using L'Hospital's rule)
\begin{equation}
\begin{array}{c}
lim \\
\lambda \rightarrow 0 \\
\end{array}%
\begin{array}{c}
\lambda ^{2}R_{ij}K^{i}K^{j} \\
\\
\end{array}%
\begin{array}{c}
=\frac{n\zeta _{0}(H_{0}-\eta _{0}X_{0})(\xi _{0}-\frac{1}{2}n(n+\gamma)\eta _{0})}{%
2X_{0}^{^{n+\gamma-1}}\left( N_{0}+\eta _{0}\sqrt{\frac{\zeta _{0}}{X_{0}^{n+\gamma-2}}}%
\right) ^{2}} \\
\end{array}%
\end{equation}%
where $H_{0}=H(X_{0},0), N_{0}=N(X_{0},0)$.\newline

The singularity is gravitationally strong in the sense of Tipler
if

\[
\xi _{0}-\frac{1}{2}n(n+\gamma)\eta _{0}>max\left\{ 0,-\frac{(n+\gamma)\Theta _{0}}{X_{0}^{\frac{%
n+\gamma}{2}}}\right\}
\]%
or
\[
\xi _{0}-\frac{1}{2}n(n+\gamma)\eta _{0}<min\left\{ 0,-\frac{(n+\gamma)\Theta _{0}}{X_{0}^{\frac{%
n+\gamma}{2}}}\right\}
\]
If the above condition is not satisfied for the values of the
parameters
then $%
\begin{array}{clll}
lim &  &  &  \\
\lambda \rightarrow 0 &  &  &
\end{array}%
\begin{array}{c}
\lambda ^{2}R_{ij}K^{i}K^{j}\leq 0{}%
\end{array}%
$ and the singularity may or may not be Tipler strong.

\section{\normalsize\bf{Discussions and Concluding Remarks}}

The paper deals with gravitational collapse in $(n+2)$ dimensional
quasi-spherical Szekeres' space-time for perfect fluid with
barotropic equation of state $p_{r}=(\gamma-1)\rho$. The role of
the parameter `$\gamma$' has been discussed in formation of naked
sigularity. It is observed that the end state of collapse (black
hole or naked singularity) remains invariant as long as the sum
`$n+\gamma$' is fixed i.e., a perfect fluid with equation of state
$p_{r}=\rho$ in `$n$' dimension, dust in $(n+1)$ dimension and
exotic matter with equation of state $p_{r}=-\rho$ in $(n+2)$
dimension are equivalent from the point of view of final state of
collapse. This interesting feature has been shown both for local
and global nature of the singularity. However, one may note that
the above results are valid for the specific choice of $C(r)$ in
the series for given in the eq.(12). Finally, the strength of the
naked singularity has been examined using the criterion
introduced by Tipler [22]. It is found that naked singularity may
be a strong curvature singularity or not depending on the choice of the parameters at $r=0$.\\\\

{\bf Acknowledgement:}\\

One of the authors (SC) is thankful to CSIR, Govt. of India for
providing a research project No. 25(0141)/05/EMR-II.\\

{\bf References:}\\
\\
$[1]$S.W. Hawking, and G.F.R. Ellis, \textit{''The Large Scale
Structure of Space-Time''.}(Cambridge University Press,
Cambridge, England. 1973).\\
$[2]$R. Penrose, \textit{Riv. Nuovo
Cim.} \textbf{1} 252 (1969); R.Penrose, \textit{In General
Relativity, an Einstein Centenary Volume, S.W. Hawking and W.
Israel (Eds.),} (Camb.
Univ.Press, Cambridge, 1979).\\
$[3]$P.S. Joshi, \textit{Pramana}
\textbf{55} 529 (2000); C. Gundlach, \textit{Living Rev. Rel.}
\textbf{2} 4 (1999); A. Krolak, \textit{Progr. Theor. Phys.
Suppl.} \textbf{136} 45 (1999); R. Penrose, \textit{In Black
Holes and Relativistic Stars, R.M. Wald (Eds),}(Univ. of Chicago
Press, Chicago,illinois, 1998); T.P. Singh \textit{gr-qc /9805066}
(1998).\\
$[4]$P.S. Joshi, \textit{Global Aspects in Gravitation and
Cosmology, }(Oxford Univ. Press, Oxford, 1993).\\
$[5]$C.J.S. Clarke, \textit{Class. Quant. Grav.} \textbf{10} 1375
(1993); T.P. Singh, \textit{J. Astrophys. Astron.} \textbf{20}
221 (1999).\\
$[6]$F.C. Mena, R. Tavakol and P.S. Joshi, \textit{Phys. Rev. D}
\textbf{62} 044001 (2000).\\
$[7]$P. S. Joshi and I. H. Dwivedi, \textit{Commun. Math. Phys.}
\textbf{166 } 117 (1994); \textit{Class. Quantum Grav.}
\textbf{16 } 41 (1999).\\
$[8]$K. Lake, \textit{Phys. Rev.Lett.}\textbf{68} 3129 (1992).\\
$[9]$A. Ori and T. Piran, \textit{Phys. Rev. Lett.} \textbf{59} 2137 (1987).\\
$[10]$T.Harada, \textit{Phys. Rev. D} \textbf{58} 104015
(1998).\\
$[11]$ A. Banerjee ,U. Debnath and S. Chakraborty, {\it
Int. J.Mod. Phys. D.} {\bf 12} 1255 (2003).\\
$[12]$U. Debnath and S. Chakraborty, {\it Gen. Rel. Grav.} {\bf 36} 1243 (2004).\\
$[13]$P. Szekeres, \textit{Phys. Rev. D} \textbf{12} 2941
(1975).\\
$[14]$ P.S. Joshi and A. Krolak, \textit{Class. Quant. Grav.}
\textbf{13} 3069 (1996).\\
$[15]$ S. S. Deshingkar ,S. Jhingan , and P. S. Joshi,
\textit{Gen. Rel.
 Grav.} \textbf{30} 1477 (1998).\\
$[16]$U. Debnath, S. Chakraborty and J. D. Barrow, {\it Gen. Rel. Grav.} {\bf 36} 231 (2004).\\
$[17]$U. Debnath and S. Chakraborty, {\it JCAP}~ {\bf 05} 001 (2004).\\
$[18]$S.Chakraborty and U.Debnath, {\it Mod. Phys. Letts. A} {\bf 20} 1451 (2005).\\
$[19]$S.Chakraborty, S.Chakraborty and U.Debnath, {\it Int. J.
Mod. Phys. D} {\bf 14} 1707 (2005).\\
$[20]$N.Dadhich, S.G Ghosh and D.W. Deshkar {\it Int. J. Mod. Phys. A} {\bf 20} 1495 (2005).\\
$[21]$P.S. Joshi and I.H. Dwivedi {\it Phys. Rev. D} {\bf 47} 5357 (1993).\\
$[22]$F.J. Tipler, {\it Phys. Lett. A.} {\bf 64} 8 (1987).\\
$[23]$C.J.S Clarke and A. Krolak {\it J. Geom. Phys.} {\bf 2} 127 (1986).\\

\end{document}